# Highly Parallel Acoustic Assembly of Microparticles into Well-Ordered Colloidal Crystallites

Crystal E. Owens,[1,2] C. Wyatt Shields IV,[1,3] Daniela F. Cruz,[1,3] Patrick Charbonneau[1,4,*] and Gabriel P. López[1,2,3,*]

1. NSF Research Triangle Materials Research Science and Engineering Center, Duke University, Durham, NC 27708, USA
2. Department of Mechanical Engineering and Materials Science, Duke University, Durham, NC 27708, USA
3. Department of Biomedical Engineering, Duke University, Durham, NC 27708, USA
4. Department of Chemistry, Duke University, Durham, NC 27708, USA

***Corresponding Authors:** *E-mail:* gabriel.lopez@duke.edu
patrick.charbonneau@duke.edu

**Abstract**

The precise arrangement of microscopic objects is critical to the development of functional materials and ornately patterned surfaces. Here, we present an acoustics-based method for the rapid arrangement of microscopic particles into organized and programmable architectures, which are periodically spaced within a square assembly chamber. This macroscale device employs two-dimensional bulk acoustic standing waves to propel particles along the base of the chamber toward pressure nodes or antinodes, depending on the acoustic contrast factor of the particle, and is capable of simultaneously creating thousands of size-limited, isotropic and anisotropic assemblies within minutes. We pair experiments with Brownian dynamics simulations to model the migration kinetics and assembly patterns of spherical microparticles. We use these insights to predict and subsequently validate the onset of buckling of the assemblies into three-dimensional clusters by experiments upon increasing the acoustic pressure amplitude and the particle concentration. The simulations are also used to inform our experiments for the assembly of non-spherical particles, which are then recovered via fluid evaporation and directly inspected by electron microscopy. This method for assembly of particles offers several notable advantages over other approaches (e.g., magnetics, electrokinetics and optical tweezing) including simplicity, speed and scalability and can also be used in concert with other such approaches for enhancing the types of assemblies achievable.

## 1. Introduction

Self-assembly enables the efficient, bottom-up generation of supra-colloidal structures that are not feasible through top-down fabrication.[1-3] The promise of this approach has inspired a wealth of research for generating structures exploiting a multiplicity of tunable interparticle interactions, including hard core (elastic),[4, 5] depletion,[6, 7] dehydration[8-10] and Coulombic interactions[11] as well as the potential to apply external fields (e.g., optical,[12] electrodynamic[13-16] or magnetic[17-19]) that couple with these interactions. A number of these methods, however, require the use of particles with narrowly defined properties (e.g., electrically polarizable or magnetically susceptible) to promote the formation of desirable structures within an external driving field.[6, 7] Passive assembly processes are materially more permissive, but are often slower, tedious and rigidly system-specific (e.g., sensitive to free charges in solution, surfactant concentrations or temperature). Perhaps more importantly, many of these assembly techniques are not scalable because of limits from field amplitudes and the complexity of particle synthesis.

We demonstrate here the use of high-frequency pressure waves to provide precise and periodic forces on microscopic particles for their acoustophoretic translation and dynamic self-assembly in fluids. The advantages of an acoustics-based approach for field-induced assembly include: (i) the nonspecific nature of the field as it interacts with a broad range of particulate matter (e.g., previous studies have exploited pressure waves to manipulate carbon nanotubes,[20] silver nanowires,[21] diamond nanoparticles,[22, 23] polystyrene spheres,[24] cell-encapsulating microscopic hydrogels,[25] aqueous droplets in oil,[26] silicone microparticles,[27] glass fibers,[28] mammalian cells[29] and unicellular organisms[30, 31]), (ii) the relative rapidity of the assembly process (i.e., on the order of seconds to minutes) , (iii) the tunability of the resulting assemblies (e.g., in size, shape and location), (iv) the precise spatial positioning of assemblies over large distances (e.g., at least several centimeters) for scalable assembly and (v) the use of a single driving element to generate a range of wavelengths, sometimes in more than one dimension.[32] The advantages for acoustics-based particle manipulation, in general, have garnered the attention of many experimentalists aiming to assemble and process synthetic and biological materials for a variety of applications. Demonstrated uses of acoustics include the fabrication of three-dimensional phononic crystals,[24] the structured infusion of particles in materials for generating composites[33] and the assembly of nanostructures,[34] as well as a plethora of biotechnological applications, including flow cytometry,[35, 36] biosensing,[37-39] cell focusing and sorting[40-44] as well as cell trapping and confinement.[27, 45, 46]

While some groups have investigated the use of standing pressure waves to arrange particles in laminar flows and multidimensional fields,[47-49] the use of acoustic radiation forces for generating well-ordered assemblies of particles has received much less attention. López, Yellen and coworkers published two studies where acoustic forces were used in the assembly of a diverse array of highly ordered crystalline colloidal structures.[50, 51] These studies demonstrated the utility of acoustic fields in tandem with magnetic fields in the generation of nontrivial particle order, but the acoustic fields utilized were not well-controlled nor well-characterized. Further, while studies on acoustic assembly are often accompanied by computational models that effectively recapitulate the structures observed empirically,[23, 24] holistic simulations capable of predicting features of typical particle trajectories and forces on (or between) particles in response to acoustic standing waves have yet to be considered.

We present a joint experimental and computational study of the rapid and highly parallel assembly of microparticles into well-ordered structures, relying solely on bulk acoustic standing waves in an assembly chamber device (Fig. 1). In Section 2, we describe a Brownian dynamics simulation capable of describing particle migration kinetics and the resulting crystallites that assemble within the device. We also present the design of the device, its operation and the methods used to measure the local pressure within the device. In Section 3, we show the programmed concentration and assembly of particles into a variety of periodic structures along the floor of the device with precise two-dimensional control. In addition to positive acoustic contrast particles, which assemble around the pressure nodes, we include the assembly of negative acoustic contrast particles, which instead cluster around the pressure antinodes, and we compare our experimental results to the predictions generated by our model and simulation. The close resemblance between experiments and simulation enabled us to extrapolate our findings to the formation of structures under unexplored operating conditions, such as the programmed buckling of two-dimensional assemblies into three-dimensional clusters at high pressure amplitudes and at high particle concentrations. Finally, we consider the assembly of anisotropically shaped particles within the acoustic field, and we analyze the resulting structures using scanning electron microscopy after evaporation of the carrier fluid.

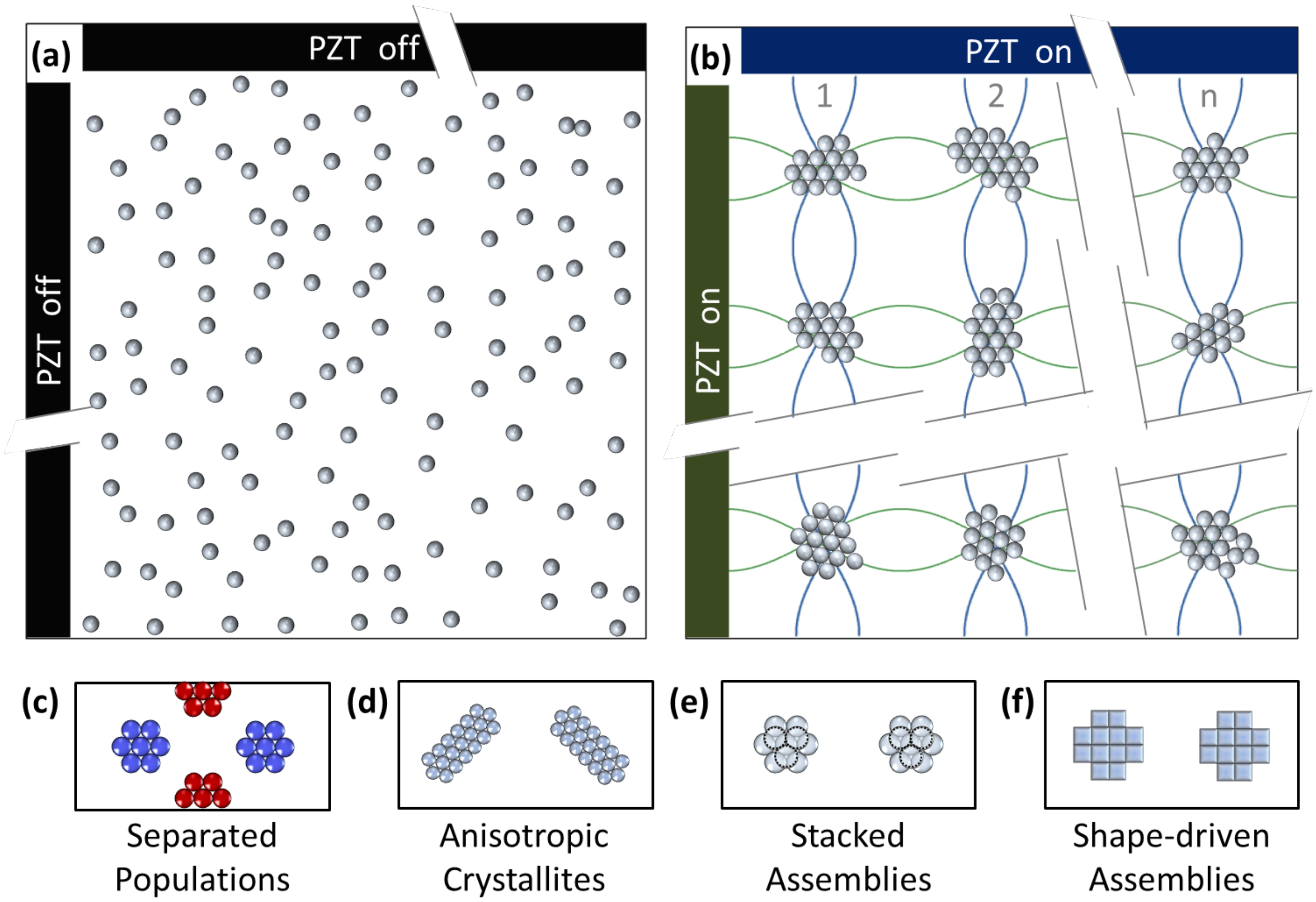

**Figure 1.** Schematic overview of particle assembly in the parallel acoustic chamber (not to scale). (a) Prior to actuation of the PZT (lead zirconate titanate) transducers, particles are randomly distributed throughout the fluid medium on the bottom (floor) of the device. (b) Following actuation of the transducers, the orthogonal standing pressure waves generate evenly spaced pressure nodes (indicated schematically by the intersection of the blue and green lines), extending over *n* half-wavelengths in each direction. Particles with positive acoustic contrast factors assemble at these nodes. The acoustic device is also capable of generating (c) demixed clusters of particles with positive and negative acoustic contrast factors, (d) anisotropic colloidal crystallites from tailoring the pressure waveforms, (e) multilayered structures from the out-of-plane buckling of two-dimensional colloidal crystallites and (f) non-hexagonally close packed colloidal crystallites from the arrangement of anisotropic particles (e.g., square cuboids).

## 2. Methods

### *2.1 Model and Brownian dynamics simulation*

Acoustophoresis is a mechanism by which pressure waves travel through a medium and exert forces on objects, driving their movement.[52, 53] We used MATLAB (MathWorks, Inc.) to implement Brownian dynamics simulations of particle acoustophoresis in a multinodal acoustic assembly chamber. We numerically integrated the time dependence of particle positions on the three-dimensional, discrete forces from primary acoustic radiation, secondary acoustic radiation, Stokes' drag, interparticle interactions and gravitation. The force equations used in this simulation are described below, with vectors noted in bold. Nomenclature is summarized in Table 1.

Particles are initially distributed randomly along the floor of the device to account for gravitational settling and are confined to a small, square section of the base of the chamber (i.e., 1.5 x 1.5 $mm^2$) under periodic boundary conditions (1.5 wavelengths in width) with a solid lower boundary and an arbitrary height. Forces are integrated over time using a Verlet scheme and a time step $\Delta t = 1 \times 10^{-6}$ sec, which is small enough for the integration to converge and large enough for the description of the Brownian motion to be reasonably accurate.[54] The number of particles, *N*, ranges from 10 to 750 in order to match the area concentrations used in experiments (once particles have settled to the base of the chamber). The particle diameter ($a$ = 1.5 to 10 μm) is much smaller than the acoustic wavelength ($\lambda$ = 1.0 mm), which falls well within the conditions of validity of the Gor'kov equation for primary acoustic radiation.[52] The potential generated by a standing acoustic wave in two dimensions on a particle much smaller than that wavelength is then[55]

$$U_{\mathrm{ac}} = \frac{V}{4\rho_m c_m^2}\left(-2\left(\frac{\beta_p}{\beta_m} - 1\right) < p(x,y)^2 > + \frac{6(-\rho_p+\rho_m)}{(2\rho_p+\rho_m)k^2} < |\nabla p(\mathrm{x},\mathrm{y})|^2 >\right), \tag{1}$$

where $V$ and $k$ are the particle volume and acoustic wavenumber, and $\rho_m, c_m, \beta_m, \rho_p, c_p$ and $\beta_p$ are the density of the medium (i.e., water; 1000 kg/m$^3$), speed of sound in the medium (1481 m/sec), compressibility of the medium (4.58x10$^{-10}$ Pa$^{-1}$), density of the particle (1050 for polystyrene and 1200 kg/m$^3$ for SU-8), speed of sound in the particle and compressibility of the particle (3x10$^{-10}$ Pa$^{-1}$ for polystyrene and 2.5x10$^{-12}$ Pa$^{-1}$ for SU-8), respectively. The acoustic pressure in the two-dimensional acoustic assembly chamber is

$$p(x,y) = p_{0x} \cos k_x x \sin \omega_x t + p_{0y} \cos k_y y \sin \omega_y t, \tag{2}$$

where $k_x, x, \omega_x$ and $p_{0x}$ are the wavenumber, position, angular frequency and pressure amplitude component due to the wave traveling in the *x*-direction, and similarly for terms with *y*-subscripts in the *y*-direction. While the pressure field is truly three-dimensional, vertical acoustic effects were not taken into consideration because the assembly occurred within a distance much less than one wavelength from the base of the chamber. We minimized the expected perturbation of particles on propagating waves by creating standing waves in a plane, which transmitted through the 3 mm height of liquid to displace particles that were <1% of that height.

The time-averaged, primary acoustic radiation force on a particle is the gradient of the potential:

$$\boldsymbol{F}_{\mathbf{PR}} = -\boldsymbol{\nabla} U_{\mathrm{ac}}. \tag{3}$$

The primary acoustic force is produced by the incident pressure wave acting on individual particles, and this force scales linearly with particle volume:[32]

$$\boldsymbol{F}_{\mathbf{PR}} = \frac{-\pi V \beta_m \phi}{2\lambda} \boldsymbol{\nabla}\langle p(x,y)^2\rangle, \tag{4}$$

where the acoustic contrast factor

$$\phi = \frac{5\rho_p - 2\rho_m}{2\rho_p + \rho_m} - \frac{\beta_p}{\beta_m} \tag{5}$$

depends on the mismatch in density and compressibility between particle and the surrounding medium. Particles with a positive acoustic contrast factor migrate to the pressure nodes of a standing wave, whereas particles with a negative acoustic contrast factor migrate to the antinodes.[27, 37, 38, 45] In two dimensions, a nonzero time-varying pressure term emerges from the interaction of two waves with unequal frequencies. This dynamical contribution was generally avoided in experiments by choosing identical frequencies for each lead zirconate titanate (PZT) piezoelectric transducer (with one exception being "Scheme IV" in Section 3.2), as its effects have been explored elsewhere.[56]

We also incorporated secondary acoustic radiation forces, which result from the scattering of pressure waves by other particles, causing higher-order interactions between particles:

$$\boldsymbol{F}_{\mathbf{SR}} = 4\pi a^6 \left[ \frac{(\rho_p - \rho_m)^2 (3cos^2\theta - 1)}{6\rho_m\, d^4} < u(x,y)^2 > - \frac{\omega^2 \rho_m (\beta_p - \beta_m)^2}{9d^2} < p(x,y)^2 > \right] \hat{\boldsymbol{e}}_{\boldsymbol{r}}, \tag{6}$$

where $d$ is the distance between a pair of particles and $\theta$ is the angle formed by the direction of acoustic wave propagation against the unit vector that connects two particles, $\hat{\boldsymbol{e}}_{\boldsymbol{r}}$.[32] We note that the fluid particle velocity $u(x,y)$ is proportional to the acoustic pressure by the fluid impedance, $\rho_m c_m$. We also note that $\boldsymbol{F}_{\mathbf{SR}}$ can either be positive or negative, depending on the angle between particles and their position in the acoustic field, and that the force is short-ranged. Hence, its effects are most significant when groups of particles are concentrated near their equilibrium positions (i.e., at the pressure nodes or antinodes).

Stokes' drag, which opposes the motion of particles through a viscous medium at low Reynolds number (i.e., within the Stokes regime), was also included:

$$\boldsymbol{F}_{\mathrm{d}} = -6\pi a\mu\boldsymbol{u}, \tag{7}$$

where $\mu$ is the viscosity of the medium ($1.002 \times 10^{-3}$ Pa·s) and $\boldsymbol{u_p}$ is the particle velocity. The colloidal interactions between particles were modeled via a truncated and shifted Lennard-Jones potential, with a cutoff radius of $\sigma = a/\sqrt[6]{2}$, to remove the attractive component of the potential (thus corresponding to the WCA potential (Weeks *et al.*) [57]):[58]

$$\boldsymbol{F}_{\mathrm{LJ}} = -12\epsilon\left[\left(\frac{a^{12}}{d^{13}}\right) - 2\left(\frac{a^{6}}{d^{7}}\right)\right]\hat{\boldsymbol{e}}_{\boldsymbol{r}}. \tag{8}$$

Since σ is the distance at which the force from colloidal interactions is zero, the truncated force does not require shifting. This interaction model was chosen because it could be integrated as a force, allowing it to be added seamlessly to the other forces in the simulation. In addition, the prefactor $\epsilon$ allows us to easily tune the energy scale for qualitatively matching the observed elasticity of collisions between particles. We found that $\epsilon = 10^{-16}$ J gives results that are close to experimental observations.

Although the effects of Brownian motion are relatively weak in our system due to the relatively large size of the particles, they were nonetheless included to account for possible thermal contributions to the assembly, including defect annealing. This stochastic process was represented by a positional update with a zero mean and normal distribution with a width of $\sqrt{6D\Delta t}$:

$$x(t + \Delta t) = x(t) + \sqrt{6D\Delta t}\, \mathcal{n}(0,1)\hat{\boldsymbol{e}}_{\boldsymbol{x}} \cdot \hat{\boldsymbol{e}}_{\boldsymbol{sx}}, \tag{9}$$

where the diffusion coefficient at room temperature, *T*, is

$$D = \frac{k_B T}{6\pi a\mu}. \tag{10}$$

The function $n(0,1)$ generates a normal distribution with mean of 0 and variance of 1, $\hat{e}_x$ is the unit vector pointing in the $x$ direction, and $\hat{e}_{sx}$ is a unit vector pointing in a random orientation in three dimensions.

We also included gravitational and buoyancy forces to recapitulate the settling of dense particles on the chamber floor and to resist out-of-plane buckling during the assembly process:

$$\boldsymbol{F_g} = (\rho_m - \rho_p)V\boldsymbol{g}. \tag{11}$$

Joining these various forces produces equations of motion that have no closed-form solution, but can provide a reasonably accurate representation of particle motion and assembly as an integrated Brownian dynamics simulation. We note that several other interactions present in our experimental system were neglected, including electrostatic interactions, van der Waals interactions and frictional drag. However, we deduce the overall contribution of these forces to be relatively weak, based both on physical arguments and on their limited impact on the motion or interaction of the particles studied herein (see Section 3 for comparisons between simulations and experiments). We show that such a simulation can provide useful predictions and illustrations for the distribution of particles over many nodes (see Video 1 in the ESI) and for the packing of particles at single-nodes into structures that can buckle out-of-plane (see Video 2 in the ESI).

**Table 1.** Nomenclature.

| **Symbol** | **Description** |
|---|---|
| $p_0$ | acoustic pressure amplitude |
| $\beta$ | compressibility |
| $\rho$ | density |
| $k$ | wavenumber = 2π/λ |
| $a$ | particle radius |
| $\lambda$ | wavelength |
| $\mu$ | viscosity of the medium |
| $\sigma$ | Lennard-Jones cutoff radius |
| $\phi$ | acoustic contrast factor |
| $\epsilon$ | depth of Lennard-Jones potential well |
| $c$ | speed of sound |
| $< p^2 >$ | mean squared pressure amplitude |
| $< u^2 >$ | mean squared fluid particle velocity amplitude |
| $d$ | distance between particles |
| $V$ | volume of the particle |
| $f$ | frequency |
| $g$ | acceleration due to gravity |
| $t$ | time |
| $u_p$ | particle velocity |
| $\theta$ | angle between line connecting particles and wave propagation direction |
| $\omega$ | angular frequency = $2\pi f$ |
| $X_p, X_m$ | subscripts 'p' and 'm' represent properties of the particle and medium, respectively |
| $t_a$ | time required for assembly |

*2.2 Device fabrication, operation and characterization*

We constructed the chamber device from similar designs used in previous studies by adhering a square chamber cut from transparent acrylic (Plexiglas, McMaster-Carr, Co.) to a 1 mm thick glass substrate (Nexterion® Glass B, Schott, AG) using a silicone adhesive (Momentive Performance Materials, Inc.).[22, 23, 59-61] The chamber (Fig. 2(a)) contained a square internal opening of 31 x 31 $mm^2$, a wall thickness of 5 mm and a wall height of 6 mm. Two PZT transducers (30 x 10 x 1 mm, 841-WFB, APC International, Ltd.) were affixed with cyanoacrylate glue (Elmer's Products, Inc.) onto two of the adjacent inner walls of the device and were oriented such that their electrical connections faced outward to prevent electrical contact with liquid inside the chamber. Two 28 x 5 $mm^2$ strips of stainless steel (0.25 mm thick, 430 annealed, Trinity Brand Industries, Inc.) were attached to the interior walls of the device opposing the PZT transducers, also using cyanoacrylate glue. These strips served as high-impedance, acoustically reflective surfaces to support the acoustic standing waves within the chamber. We designed and operated our system to support 3,600 nodes (30 x 30 wavelengths of a 1.0 mm acoustic standing wave) or 256 nodes (8 x 8 wavelengths of a 3.75 mm acoustic standing wave) when assembling large particles (i.e., > 20 μm).

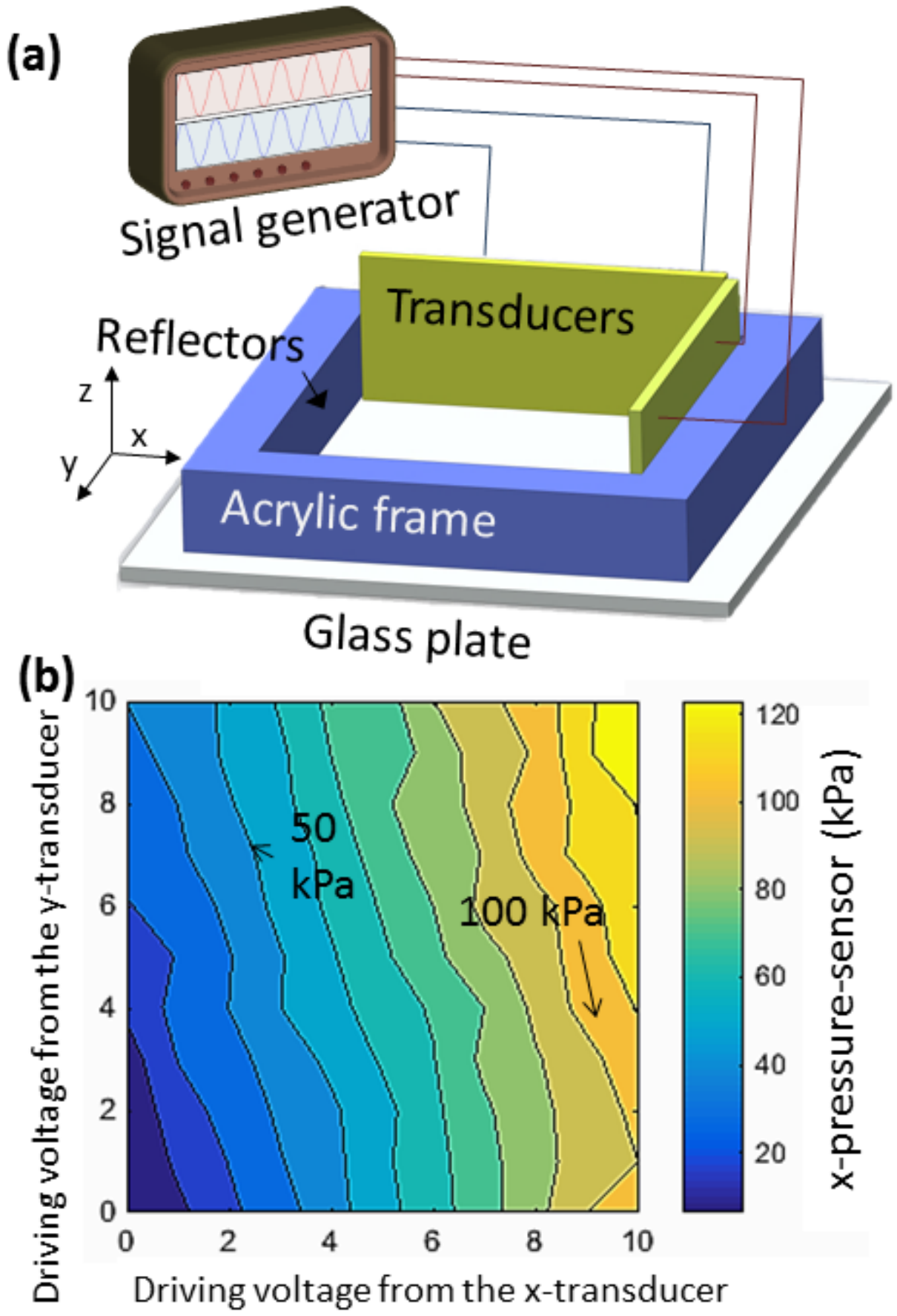

**Figure 2.** Design and operational pressures of the acoustic assembly chamber. (a) The chamber is comprised of an acrylic frame adhered to a glass plate, driven by two PZT transducers on adjacent walls. (b) Pressure amplitudes ($x$-pressure sensor) were measured by a calibrated, sensing PZT transducer opposite of the driving PZT transducer in the $x$-direction when both driving PZT transducers were actuated at a variety of voltage inputs at a constant frequency of 3.064 MHz (see the ESI for details on the measurements). The total measured acoustic pressure arises from waves emanating from both $x$- and $y$-transducers. The curves show a total operating pressure amplitude range of 0 to 120 kPa for operating voltages of 10 V peak-to-peak or less.

### *2.3 Pressure amplitude determination, sample preparation and data acquisition*

To obtain the necessary input parameters for our simulation, we measured the pressure of the standing wave inside the chamber with a standard calibration approach using a hydrophone and a sensing PZT transducer. The sensing transducer was identical to the driving transducers, and hence it was easily integrated into an operational chamber. To calibrate the transducer for pressure, we first submerged a driving PZT transducer in a large water bath and measured the amplitudes of the outgoing pressure waves with a hydrophone fixed at an arbitrary distance of 25 mm. We then measured the amplitudes of the outgoing pressure waves with a sensing PZT transducer also fixed at 25 mm to generate a voltage-pressure calibration curve for a resonant driving frequency of the chamber at AC voltages ranging from 0 to 10 V peak-to-peak (see Fig. S1 in the ESI for the experimental setup and the resulting calibration curve). We chose a driving frequency of 3.064 MHz to generate a standing wave 30 wavelengths long. This frequency delivered larger potential pressure amplitudes than at other frequencies as it was near the resonant frequency of the PZT element (2.9 MHz). Once the calibration was established, we placed the sensing PZT transducer inside the chamber device along the wall opposite to the driving PZT transducer in the *x*-direction, and we used it to determine the pressure amplitude emanating from both PZT transducers. As expected, we observed a coupling of vibration in the walls of the chamber (i.e., along the *x*- and *y*-directions) when excited by a single transducer, as indicated non-zero values of pressure measured by the sensing transducer along the *x*-direction from the actuation of the driving transducer along the *y*-direction (Fig. 2b). This coupling occurred because the square geometry of the chamber allows resonance to occur simultaneously in the *x*- and *y*-directions at a single frequency. The pressure amplitudes generated in our device range from tens to over one hundred kilopascals, which is in agreement with previous determinations for similar devices.[23, 24]

We prepared suspensions of positive acoustic contrast particles by suspending fluorescent polystyrene spheres (10.8 µm diameter, Spherotech, Inc.) in 0.1 vol.% Tween 20 (Sigma Aldrich, Co.) in deionized water at concentrations from $5 \times 10^4$ to $5 \times 10^6$ particles/mL. We then added 3 mL of the particle suspension to the chamber, which produced a fluid layer about 3 mm deep with a slight downward meniscus. Particle sedimentation resulted in area concentrations ranging from 50 to 1,000 particles/$mm^2$ on the floor of the chamber. To add negative acoustic contrast particles to the suspension, we replaced 1 mL of deionized water with a suspension of red fluorescent silicone spheres (1.5 µm diameter) synthesized from a stock solution with a 96:1 molar ratio of dimethoxymethylsilane:tetramethoxysilane (Sigma Aldrich, Co.) using a method described previously.[27] We prepared suspensions of anisotropically shaped, positive acoustic contrast particles (shown in Section 3.5) by projection photolithography, using a method also described previously,[16, 62] to make 50 x 10 µm square, hexagonal and circular prisms, and we suspended these particles in a solution of 0.1 vol.% Tween 20 in deionized water.

A signal generator (AFG3022C, Tektronix, Inc.) applied a continuous sinusoidal wave of voltages (1-10 V peak-to-peak) to the PZT transducers, where the signals were in phase between the PZT transducers. For experiments requiring high pressure amplitudes (e.g., $p_0$ > 120 kPa), we used a power amplifier (25A250AM6, Amplifier Research Corp.) in series with the signal generator to increase the voltage supplied to the PZT transducers. We monitored the output from the amplifier with an oscilloscope (TDS2004C, Tektronix, Inc.) and converted the values to units of pressure using the calibration approach described earlier. Particles were imaged throughout the assembly process using a fluorescence microscope (Axio Imager 2, Carl Zeiss, AG) or a confocal microscope (780 upright confocal, Carl Zeiss, AG). Following overnight evaporation of the fluid within the chamber (with the PZT transducers on), we decoupled the chamber from the substrate using a razor blade; we then deposited approximately 8 nm of gold (Desk IV sputter coater, Denton Vacuum, Inc.) onto the substrate and inspected the assemblies with a scanning electron microscope (SEM; XL30 ESEM, FEI Co.).

## 3. Results and Discussion

### *3.1 Model and simulation of the acoustic assembly chamber*

The physics underlying the exertion of force from ultrasound on microscopic particles in suspension is well-understood and is integrated into our simulation.[47, 48] The simulated assembly could thus be directly compared to the experimental results. We separated a mixture of 10 μm cubes with positive acoustic contrast and 1.5 μm spheres with negative acoustic contrast (Fig. 3). For the sake of simplicity, we approximated the 10 μm cubes as spheres in our simulation. Yet, the close match between the two approaches demonstrates both the reasonableness of the model and the high degree of experimental control. These simulations, which ignore forces from acoustic streaming or from other sources (e.g., acoustic beating[56, 63]), reveal that the experiments behave nearly ideally for times $t < 5$ min. In addition, the microparticles used in this study diffuse slowly, so they do not undergo significant displacement upon or after assembly. We indeed found that particles were driven to the energy minima of the acoustic field, and we can thus predict the final locations of the structures from the acoustic energy field alone (Eqn. 1).

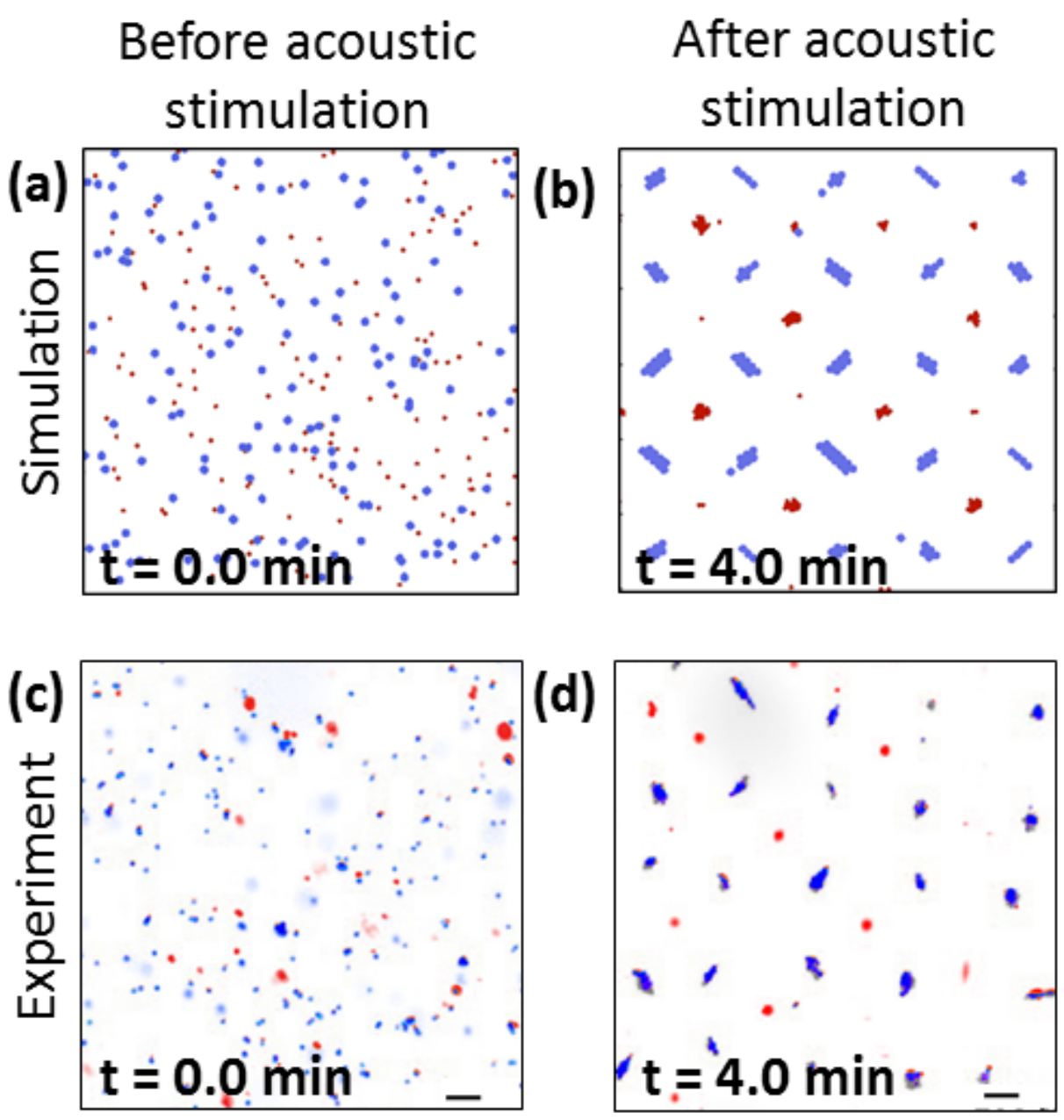

**Figure 3.** Examples of acoustically induced concentration of particles from simulation (top) and experiments (bottom). (a, c) Prior to activating the acoustic field, positive acoustic contrast particles (blue) and negative acoustic contrast particles (red) are randomly distributed along the floor of the chamber. (b, d) Orthogonal acoustic standing waves separate particles to the pressure nodes and antinodes, depending on their acoustic contrast factors. The process takes place over tens of seconds, where the final formations are shown at $t$ = 4 min. A time-lapse video of this process for different particles is provided in the ESI (see Video 3). The 10 um cubes shown in the experiment (c, d) are approximated as 10 um spheres in the simulation (a, b). We performed a background subtraction on images in (c) and (d) to improve visibility of the particles. Scale bars are 100 µm.

*3.2 Emergence of tunable nodal and antinodal patterns*

Simulations suggested that altering the ratio of forces exerted by the PZT transducers, $P_{X/Y} = (p_{0x}/p_{0y})^2$, and the difference in frequency of the signal supplied to the two transducers, $|f_X$

- $f_Y$|, can give rise to a rich array of shapes of the energy maxima and minima. We explored four specific assembly schemes, or combinations of these parameters in detail (Schemes I-IV are described in Table 2). (See Fig. S2 in the ESI for further exploration of the parameter space.) In Schemes I-III, we tuned the anisotropy of the field shape by varying $P_{X/Y}$, and in Scheme IV we added a small frequency difference to generate two-dimensional acoustic beating in the field.

We performed both simulations and experiments of these schemes for a binary mixture of particles with opposing acoustic contrast factors. The results agree to a large extent, so the following discussion pertains to both simulation and experiment, except when noted. We also include fast Fourier transforms (FFTs) of the assembly patterns produced by the simulation to highlight the periodicity between clusters formed (i) solely at the pressure nodes, (ii) solely at the pressure antinodes and (iii) both at the pressure nodes and antinodes (see Fig. S3 in the ESI).

**Table 2.** Description of parameters for each assembly scheme.

| Scheme | Description | Description |
|---|---|---|
| I | A single PZT transducer is activated | $P_{X/Y} = (p_{ox}/p_{oy})^2 = 0$ |
| II | The second transducer is supplied a signal with lower voltage amplitudes than the first, with equal frequency | $P_{X/Y} = 0.5$<br>$f_X - f_Y = 0$ |
| III | Transducers are supplied with voltage signals of the same amplitude and frequency | $P_{X/Y} = 1$<br>$f_X - f_Y = 0$ |
| IV | The transducers are supplied with equal voltage amplitudes, but a small difference in frequency | $P_{X/Y} = 1$<br>$\|f_X - f_Y\| > 0$ |

As expected, the acoustic assembly chamber rapidly separated the two types of particles at the pressure nodes and antinodes for Schemes I-IV. In Scheme I, both types of particles assembled along lines parallel to the active PZT transducer. Particles with positive and negative acoustic contrast factors, however, separated into the pressure nodes and antinodes, respectively, forming two linear arrays offset by a quarter of a wavelength, $\lambda/4$ = 250 µm (Fig. 4(i)). A video of positive acoustic contrast particles forming into Scheme I and III assemblies in sequence is included (see Video 4 in the ESI). We note that beading occurred within particle chains over time in experiments, where sections of the lines pulled together. A similar effect has also been observed in magnetic self-assembly, where secondary magnetic interactions perturb the overall linear assembly over time.[64] Here, we ascribe this process to secondary acoustic radiation forces, which are attractive between particles arranged orthogonal to the propagating wave (see Eqn. 6), and to coupling of vibrations in orthogonal directions due to chamber design (see Section 2.3).

In Scheme II, positive acoustic contrast particles formed periodic strings parallel to the dominant transducer in a manner similar to Scheme I, but the strings were modulated, giving a serpentine morphology to the assembly (Fig. 4(j)), as previously reported.[56] Negative acoustic contrast particles, however, contracted into periodically-spaced ellipsoids. Our model suggests that this difference is due to the stronger coupling of the negative acoustic contrast particles with the pressure field, which is likely due to the higher magnitude (or absolute value) of their acoustic contrast factor (i.e., $\phi \approx -0.88$ for silicone vs. $\phi \approx +0.46$ for polystyrene). As such, the energy wells around pressure nodes are less steep than wells around pressure antinodes, making them more susceptible to programmed variations in shape by the driving waves. Thus, positive acoustic contrast particles formed linear, oblique, curved and isotropic structures while the negative acoustic contrast particles typically formed isotropic islands or lines.

Fully operating both orthogonal fields in Scheme III led to the formation a square lattice of nodal and antinodal patterns (Fig. 4(k)). It contained twice as many antinode positions as in Scheme II. After long periods of time (i.e., $t > 5$ min), clusters of particles began to drift loosely around their expected nodal position. This behavior is characteristic of the influence of delayed effects observed after long acoustic activation times, as has been reported in other planar acoustic devices.[65]

In Scheme IV, we found that positive acoustic contrast particles formed diagonally elongated oval patterns with alternating orientations in the pressure nodes, forming a chiral pattern (Fig. 4(l)). We found that a time-varying term in the two-dimensional pressure field emerged in this scheme, which is likely due to acoustic beating and occurs when pressure waves at slightly different frequencies (e.g., 0.1 to 10 Hz difference) interfere. Others have previously reported this phenomenon, and it may be the cause of the periodic motion of the particles at the nodes within the pressure field observed after long periods of time (e.g., > 5 min).[56, 63] Minor acoustic beating was present in Schemes II and III, consistent with the low level of noise produced within the experimental system. However, this effect had no impact on structures formed in $t < 5$ min. Other contributions to later perturbations (i.e., at $t > 5$min) may include an altered resonant frequency of the assembly chamber over time due to particle aggregation,[66] thermal effects (e.g., due to the PZT elements heating the fluid) and fluid convection.[44]

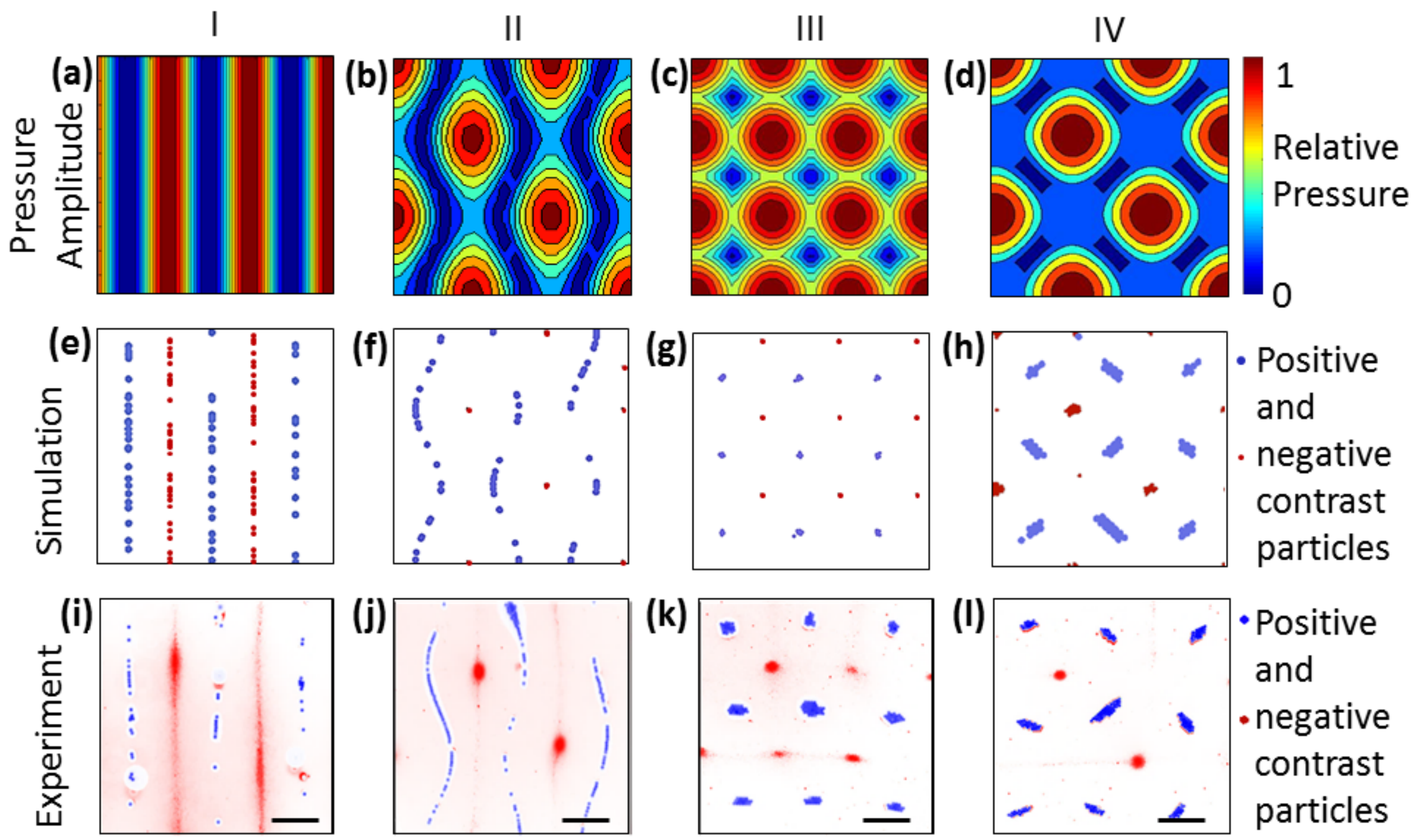

**Figure 4.** Representative mesoscale arrangement of particles within the acoustic chamber device. Four distinct assembly schemes (I-IV) are generated by varying the frequency and voltage supplied to each PZT transducer (see Table 2). For each of the four schemes, we show a square section that is 1.5 wavelengths across of: (a-d) the calculated relative acoustic standing wave landscapes from low (blue, representing the pressure nodes) to high (red, representing the pressure antinodes), (e-h) simulations of steady-state patterns of particles after imposing acoustic standing waves and (i-l) optical micrographs of the assemblies resulting from mixtures of positive and negative acoustic contrast particles at $t \approx 4$ min. The particles simulated in (e-h) match the physical characteristics of the particles shown in (i-l). Positive acoustic contrast particles are polystyrene (10.8 µm diameter) and negative acoustic contrast particles are silicone (1.5 µm diameter). We performed a background subtraction on images in panels (i-l) to improve visibility of the particles. Scale bars are 250 µm.

*3.3 Kinetics of particle assembly*

Since the entire assembly process occurs within a convenient experimental timescale (e.g., within seconds to a few minutes), direct comparison with our simulations can be used to gain insights into the detailed kinetic phenomena at work during the assembly process (Fig. 5(a-b)). We calculated each force component described in Section 2.1, and averaged over particles during the course of the simulation (Fig. 5(c)). This analysis would be impossible by experimental observation alone, but can be useful when manipulating particles (e.g., cells) that may be sensitive to large pressure amplitudes and shear. Information gained from these simulations may thus suggest modifications to the operation or geometry of the device to mitigate such effects. The results in Figure 5c indeed indicate that the primary radiation force acting on the particles decreases as particles approaches the energy minima, which corroborates with the shape of the pressure amplitude (see Fig. 4).

We studied the kinetic behavior (i.e., acoustophoresis) of polystyrene spheres by simulation and experiment (with $P_{X/Y}=1; f_X - f_Y = 0$), and we identified four distinct dynamical regimes during their assembly process: I) concentration, when particles begin to approach the pressure nodes; II) organization, when the particles begin to assemble (and the primary acoustic force begins to decrease); III) retention, when a close-packed arrangement prevents further particle motion; and IV) dissociation, when the acoustic field was removed (Fig. 5(d)). We found it particularly useful to identify the second regime because it is in that regime that particle-particle interactions became increasingly prevalent.

We defined a parameter for the time of assembly, $t_a$, which is arbitrarily defined as the time to reach the start of Regime II, to establish quantitative correlations between simulation and experiment. Simulations were used to predict $t_a$ for an arbitrary initial concentration of polystyrene beads (i.e., 300 particles/mm$^2$) at a variety of pressure amplitudes (Fig. 5(e)). We compared the predictions of $t_a$ from simulation with experimentally assembled structures, and we found that, in both simulation and experiment, $t_a \approx 5$ min for 60 kPa and $t_a < 1$ min for pressures above 100 kPa (see inset of (Fig. 5(e)). These results indicate that simulations are useful not only for predicting the equilibrium arrangement of particles in the acoustic chamber, but also for modeling the assembly kinetics. It is thus reasonable to assume that the model and simulation can also be applied to devices with different dimensions and boundary conditions as well as to particles of different sizes, shapes or physical properties (e.g., density or stiffness).

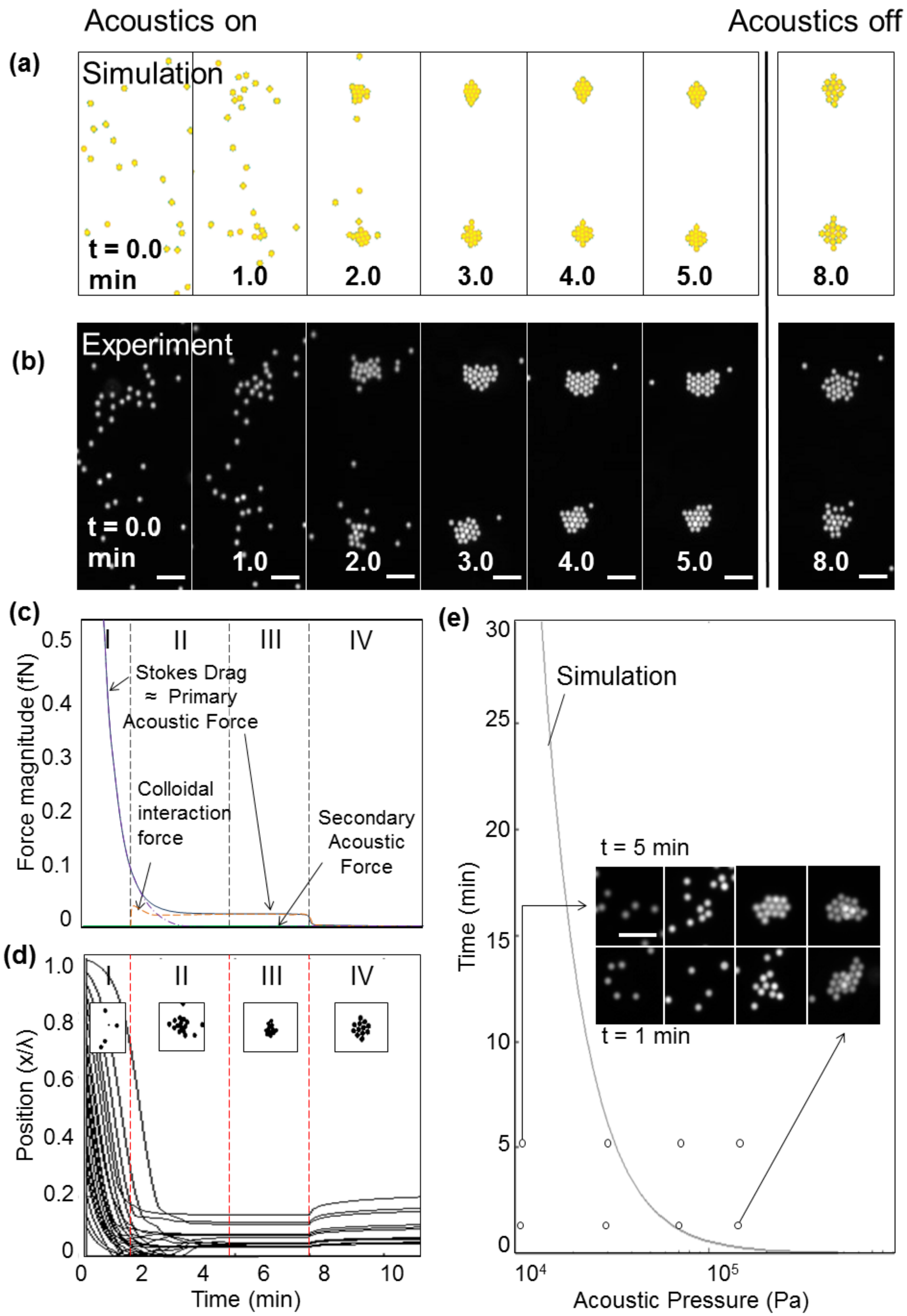

**Figure 5.** Dynamics of the assembly of particles with a positive acoustic contrast factor. (a) Simulation frames and (b) microscope images of 10 µm polystyrene spheres at the same time points following acoustic actuation (with $p_o$= 50 kPa and at $t$ = 0 min). (c) Magnitude of the major forces involved in particle assembly, averaged across each particle in the simulation. The results are subdivided into four approximate regimes: I) concentration, II) organization, III) retention and IV) dissociation. (d) Normalized distance, $x/\lambda_x$, of 10 particles from the analytically determined pressure node (micrographs are shown as insets). (e) Time required for a random suspension of 10 µm polystyrene beads to enter Regime II ($t_a$) for different acoustic pressure amplitudes, simulation and experimental results overlaid with matching experimental images. Scale bars are 50 µm.

*3.4 Programmable buckling of colloidal crystallites into stacked clusters*

We further experimentally explored the acoustophoretic translation of 10 µm polystyrene particles in response to pressure inputs ranging from 10 to 120 kPa (i.e. $P_{X/Y}$ = 1; $f_X - f_Y = 0$). As expected, at low pressures (e.g., ≈10 kPa), the particles did not appreciably concentrate (see Regime I in Fig. 5) after 10 min (Row I in Fig. 6). At slightly higher pressures (e.g., ≈30 kPa), the particles began to concentrate after 10 min (Row II in Fig. 6); however, they did not form crystallites (see Regime II in Fig. 5). At intermediate pressures (e.g., ≈70 kPa), the particles concentrated at around 2.5 min, assembled at around 5 min and stabilized (see Regime III in Fig. 5) at around 10 min (Row III in Fig. 6). Interestingly, at high pressures (e.g., ≈120 kPa), the particles rapidly organized and began to buckle out-of-plane after 1 min (Row IV in Fig. 6).

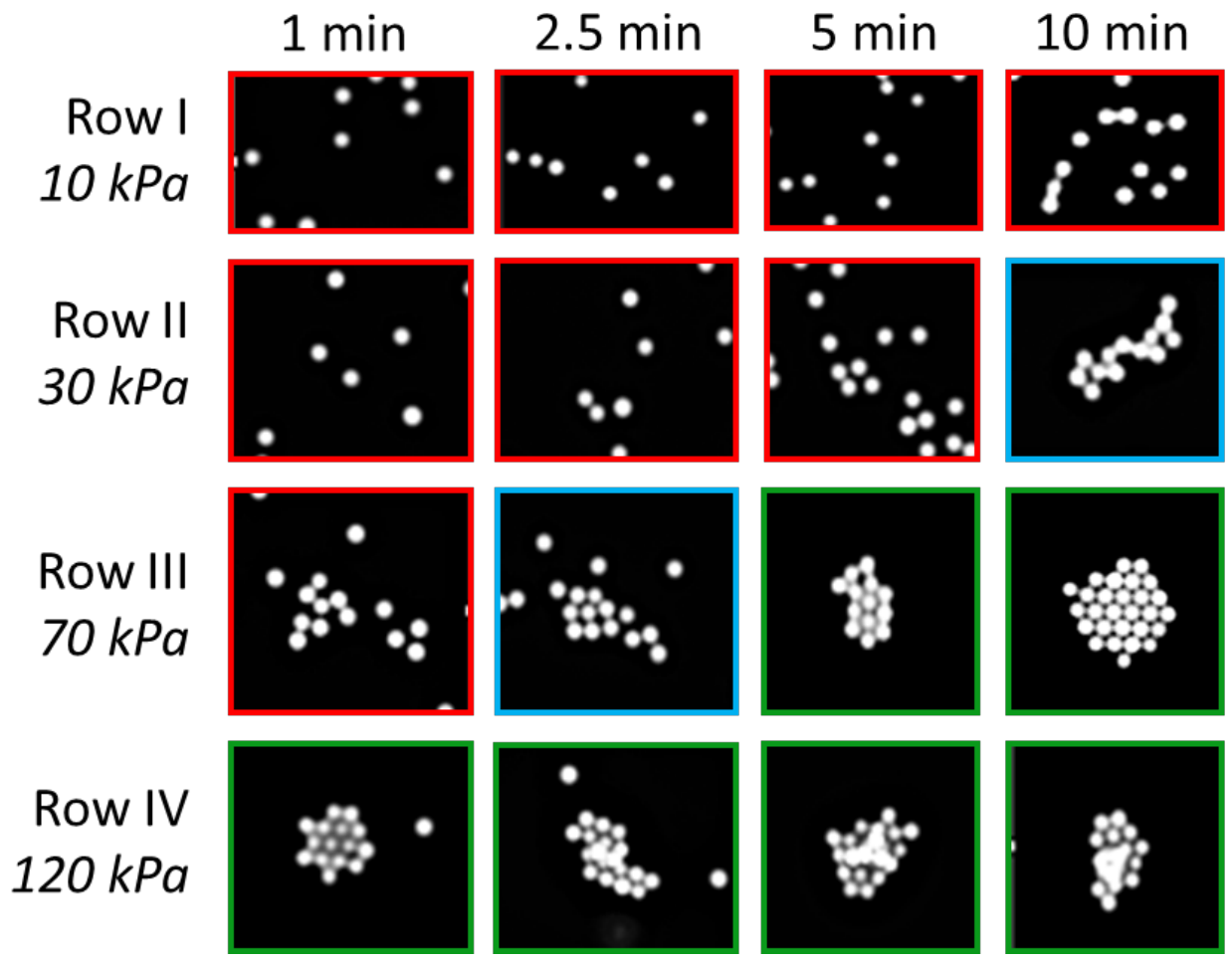

**Figure 6.** Concentration, assembly and out-of-plane buckling of spherical particles. Fluorescence microscopy images of 10 µm polystyrene particles migrating in response to an acoustic standing wave at various pressures (10-120 kPa) were collected over a 10 min period. The colored box around each image represents different dynamic regimes of particle manipulation: red for concentration, blue for organization and green for retention and out-of-plane buckling. Note: images within a row may not necessarily be from the same experiment.

Gleaning insights from these experiments, we used simulations to predict the buckling of the colloidal crystallites at high acoustic pressure amplitudes (e.g., $p_0 > \approx 75$ kPa) and at high initial particle concentrations (e.g., $C_0 > \approx 750$ particles/mm$^2$ or $\approx 2\text{x}10^5$ particles/mL) (see Video 2 in the ESI). The predictions for multilayered assembly (red filled symbols in Fig. 7(a) and Fig. 7(b-d)) were then tested by conducting experiments at a range of pressure amplitudes and initial particle concentrations (navy filled symbols in Fig. 7(a)). We identified the emergence of three-dimensional structures by epifluorescence microscopy (Fig. 7(e,f)) and by confocal microscopy (Fig. 7(g)). The simulation successfully predicted the formation of monolayered and stacked (multilayered) structures across a wide range of acoustic pressure amplitudes (i.e., 20 to 500 kPa) and initial particle concentrations (i.e., 50 to 750 particles/mm$^2$).

We also used simulations to explore conditions beyond the operating range of our chamber to predict the formation of structures much taller than the bilayer structures typically observed (see Fig. 7(d)). By amplifying the applied voltage to generate a pressure amplitude of approximately 500 kPa, which is near the maximum pressure amplitude achievable within our device, we were able to create and image a structure of stacked particles approximately 35 μm tall, as confirmed by confocal microscopy (Fig. 7(g)). This result suggests that it may be possible to generate pyramidal or even columnar structures at even higher acoustic pressures. Although our chamber device cannot achieve the output pressures necessary to assemble these structures, we hypothesized that it may also be possible to achieve structures of a similar height at lower pressure amplitudes by changing the physical properties of the particles and the medium (e.g., their density and compressibility) as to enhance their acoustic contrast and correspondingly increase the magnitude of applied radiation forces. We note, however, that the field conditions necessary to generate these assemblies in practice may result in unanticipated thermal or scattering effects that are not included in our simulation and could alter their morphology.[67, 68]

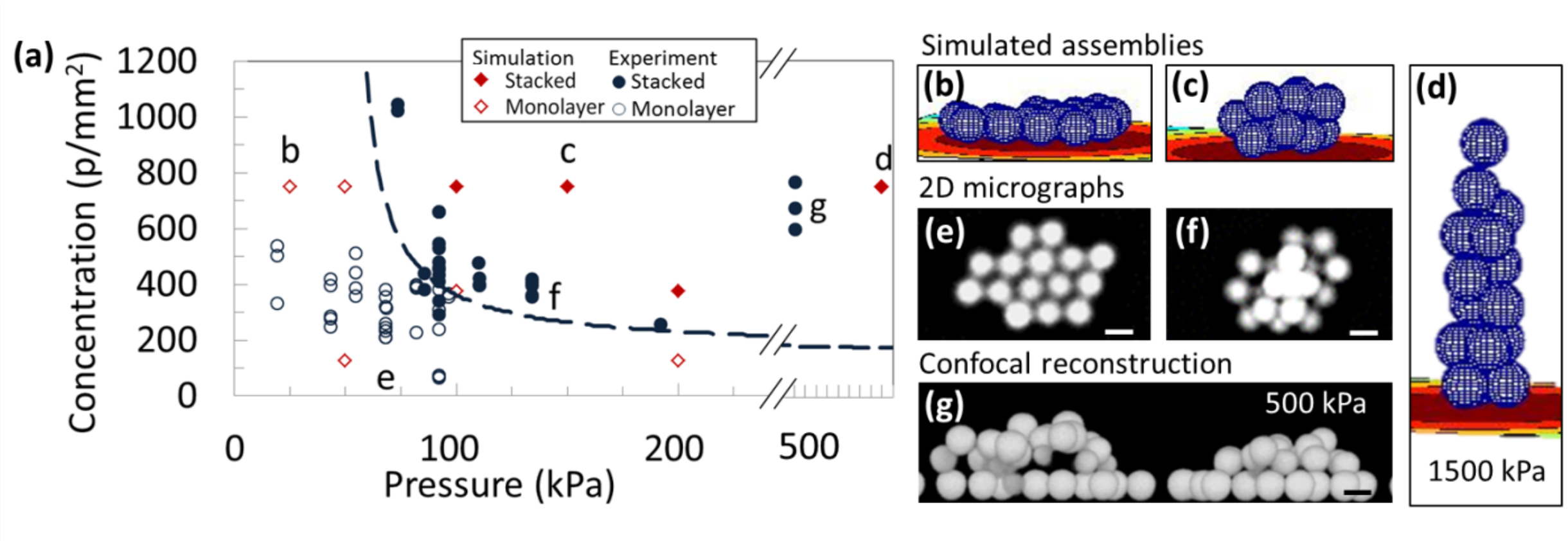

**Figure 7.** Particle stacking as a function of initial particle concentration and pressure amplitude. (a) The data points show monolayered (unfilled symbols) and multilayered (filled symbols) structures based on data from a series of simulations (red diamonds) and experiments (blue circles). The dashed line indicates the approximate transition between conditions the form monolayered versus multilayered structures as predicted by an energy model (see S5 for the fitting equation). Representative images from simulation (b, c and d) and experiments (e, f and g) are shown. Note that (d) and (g) occur at much higher pressures than those shown on the plot (i.e., 1500 and 500 kPa, respectively). Scale bars are 10 μm.

*3.5 Assembly of shaped particles into well-defined colloidal crystallites at the pressure nodes*

We further compared the two-dimensional assembly ($P_{X/Y} = 1; f_X - f_Y = 0$) of spherical polystyrene particles and non-spherical particles (i.e., 50 x 10 μm square, circular and hexagonal tiles) made from SU-8 photoresist and having a positive acoustic contrast factor in water ($\phi \approx +1.17$). Like the polystyrene beads, SU-8 particles migrated to the pressure nodes of the standing wave. After removing the acoustic force, the assemblies retained their shape due to their large size and correspondingly low diffusivity. To prepare the samples for high-resolution imaging, for particles of a relatively low material density (e.g., polystyrene), the assemblies were subsequently dried with the acoustic field on, whereas the acoustic field was removed for particles of a relatively high material density (e.g., SU-8). After drying, we decoupled the walls of the device from the glass substrate using a razor blade. This facile deconstruction allowed us to sputter coat the top of the colloidal crystallites with a thin layer of gold and then image them using a scanning electron microscope (Fig. 8).

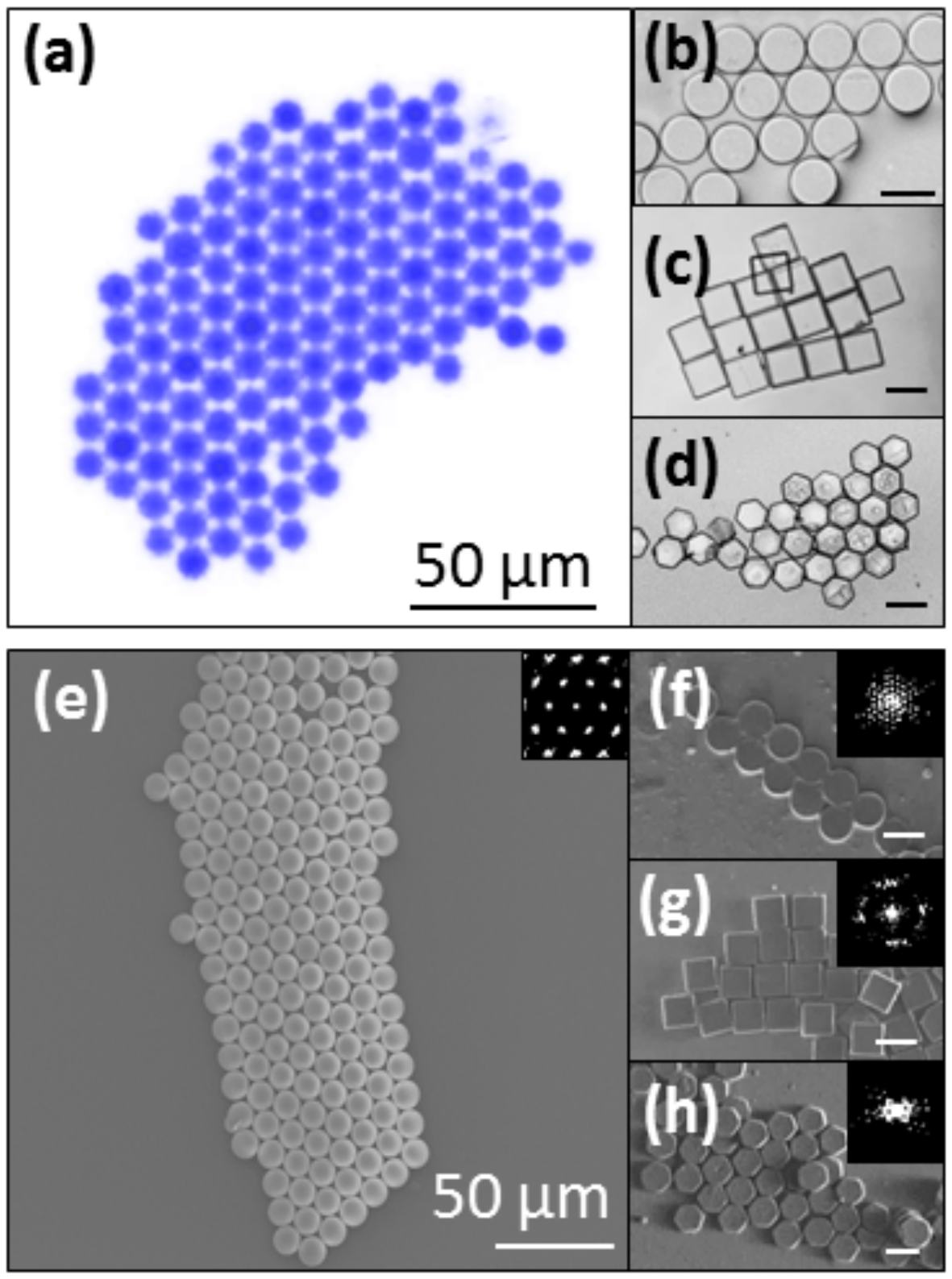

**Figure 8.** Representative micrographs of acoustically assembled colloidal crystallites. (a) Fluorescent and (e) electron micrographs of 10 µm spheres assembled into a hexagonally close packed arrangement. (b) Optical and (f) electron micrographs of 50 x 10 µm circular tiles assembled into a hexagonally close packed arrangement. (c) Optical and (g) electron micrographs of 50 x 10 µm square tiles assembled into a square close packed arrangement. (d) Optical and (h) electron micrographs of 50 x 10 µm hexagonal tiles assembled into a hexagonally close packed arrangement. The insets (top right) show the fast Fourier transforms of the dried structures for images (e-h), which indicate the particles arranged within the assemblies display some crystallinity. We performed a background subtraction on (a) to improve visibility of the particles while still in solution. Scale bars are 50 µm.

These assemblies were obtained by imposing an acoustic standing wave eight wavelengths across in both the *x*- and *y*-directions to increase the width of the potential energy well ($f_0$ = 0.75 MHz) to accommodate the larger particles. A 20x amplifier was required to increase the maximum pressure amplitude to a value high enough for particles to assemble into close packed arrangements ($P_0$ > 250 kPa). The size of the colloidal crystallites was easily tunable from tens to hundreds of particles, depending on the size of the energy well generated and the initial particle concentration The resultant dried crystallites were located at the nodal points, as expected. We note that comparable structures could not be found in dried solutions that had not been acoustically focused beforehand. We can thus confidently attribute the formation of these structures to acoustic perturbations and not to evaporation processes, such as the coffee ring effect.[69]

As expected from previous bulk simulation studies,[70] we found that particle shape directly impacts the packing arrangements. The assembly of spherical particles (Fig. 8(a, e)), circular tiles (Fig. 8(b, f)) and hexagonal tiles (Fig. 8(d, h)) formed hexagonally close packed crystallites, whereas square-shaped particles formed crystallites with an open, square symmetry (Fig. 8(c, g)). Unlike the colloidal crystallites formed from polystyrene spheres, closed structures formed from the SU-8 tile particles displayed a flat top surface morphology. This may have utility for applications, such as cell anchoring and programmed differentiation.[71] Assemblies formed from square and hexagonal tiles contained few interstitial holes (or sometimes thin lines between rows of particles), unlike the assemblies of spherical particles or circular tiles. This effect could have useful implications for applications in colloidal lithography, wherein crystalline monolayers are used as templates to direct the deposition of various materials (e.g., metals, polymers or biological reagents) for well-defined surface patterning.[72]

## 4. Conclusions

We have developed a noncontact system to rapidly assemble colloidal microparticles in a highly programmable and parallel fashion. We show four patterns, as examples, achieved from the assembly of spherical microparticles through the controlled interference of two orthogonal plane waves and explored the directed assembly of two-dimensional colloidal crystallites and their buckling into three-dimensional structures at individual pressure nodes. In tandem with this assembly approach, we used Brownian dynamics simulations to predict the kinetics of particle migration and to explore the outcomes of the various assembly schemes.

Simulation and theory were instrumental in efficiently exploring the range of system parameters leading to global assembly patterns at the pressure nodes and antinodes. They revealed the experimental conditions necessary to form stacked assemblies (i.e., those formed from the buckling of two-dimensional colloidal crystallites out-of-plane) and also predicted the formation of structures beyond the direct capabilities of our particular device. The high degree of correspondence between simulation and experiment further validated our understanding of the underlying physical mechanisms for assembly. Although this simulation includes forces from several sources, we note that the forces due to primary acoustic radiation, Stokes drag and volume exclusion dominate in this system at the timescales considered. This effect explains why a simplified description of particle interactions captures the experimental assembly process with such accuracy. For simulating the behavior of particles much smaller than those shown here with comparable contrast factors, other forces (e.g., from electrostatic and van der Waals interactions) should also be considered. We were also able to recover the assembled microstructures by simple dehydration for inspection by electron microscopy, which could also enable other downstream applications, such as templating for colloidal lithography or biological surface patterning. In short, our device enables a rapid and facile approach to generate large quantities of assembled structures in parallel in a directly accessible format. In addition, the schemes shown here can be translated simply to more complex devices, such as those with an extensive matrix of driving elements, to facilitate more exquisitely tailored patterns.[73]

The examples of the patterns presented and analyzed and should enable a variety of future studies. For example, using insights gained from this model and simulation, one can tune the elongation of the pressure node points or alter the linearity of the resulting structures. The combination of well-understood acoustic chambers with other external fields (e.g., magnetic[50, 51]) has the potential to greatly expand the types of single and multicomponent colloidal assemblies possible. The methods shown in this study thus hold significant promise for a range of potential applications in materials science and engineering, including the development of metamaterials,[22] micromirrors,[74] templates for colloidal lithography[71] and biosensors.[75] Beyond colloidal assembly studies, the rapid separation phenomena of particles with opposing acoustic contrast factors also shows potential for biological applications, such as in separating cells labeled with negative acoustic contrast particles from unlabeled cells[37, 45] for downstream analysis.[76] This method could also be used to concentrate cells for culture and tissue engineering.[25, 77]

## Acknowledgements

This work was supported by the National Science Foundation's (NSF's) Research Triangle Materials Research Science and Engineering Center (MRSEC) under Grant No. (DMR-1121107), an NSF grant from the Nanomanufacturing Program (CMMI-1363483), NSF Graduate Research Fellowships (GRF-1106401) to C.W.S. and D.F.C., and a Duke University

Alstadt SUUP Fellowship to C.E.O. We gratefully acknowledge Heath Martin and Paul Dayton at the University of North Carolina, Chapel Hill for assistance in pressure measurements and Lin Fu at Duke University for helpful discussions.

**References**

1. V. N. Manoharan, *Science*, 2015, **349**, 1253751.
2. B. Bharti and O. D. Velev, *Langmuir*, 2015, **31**, 7897-7908.
3. G. M. Whitesides and B. Grzybowski, *Science*, 2002, **295**, 2418-2421.
4. A. Haji-Akbari, M. Engel, A. S. Keys, X. Zheng, R. G. Petschek, P. Palffy-Muhoray and S. C. Glotzer, *Nature*, 2009, **462**, 773-777.
5. P. F. Damasceno, M. Engel and S. C. Glotzer, *Science*, 2012, **337**, 453-457.
6. Y. Wang, Y. Wang, X. Zheng, G. R. Yi, S. Sacanna, D. J. Pine and M. Weck, *Journal of the American Chemical Society*, 2014, **136**, 6866-6869.
7. S. Sacanna, M. Korpics, K. Rodriguez, L. Colon-Melendez, S. H. Kim, D. J. Pine and G. R. Yi, *Nature communications*, 2013, **4**, 1688.
8. N. D. Denkov, O. D. Velev, P. A. Kralchevsky, I. B. Ivanov, H. Yoshimura and K. Nagayama, *Nature*, 1993, **361**, 26.
9. G. Singh, S. Pillai, A. Arpanaei and P. Kingshott, *Soft Matter*, 2011, **7**, 3290-3294.
10. G. Singh, H. J. Griesser, K. Bremmell and P. Kingshott, *Advanced Materials*, 2011, **21**, 540-546.
11. M. E. Leunissen, C. G. Christova, A. P. Hynninen, C. P. Royall, A. I. Campbell, A. Imhof, M. Dijkstra, R. van Roij and A. van Blaaderen, *Nature*, 2005, **437**, 235-240.
12. T. Čižmár, L. C. D. Romero, K. Dholakia and D. L. Andrews, *Journal of Physics B: Atomic, Molecular and Optical Physics*, 2010, **43**, 102001.
13. S. Gangwal, A. Pawar, I. Kretzschmar and O. D. Velev, *Soft Matter*, 2010, **6**, 1413.
14. O. D. Velev and K. H. Bhatt, *Soft Matter*, 2006, **2**, 738.
15. S. Gangwal, O. Cayre, M. Bazant and O. Velev, *Physical Review Letters*, 2008, **100**.
16. C. W. Shields IV, S. Zhu, Y. Yang, B. Bharti, J. Liu, B. B. Yellen, O. D. Velev and G. P. López, *Soft Matter*, 2013, **9**, 9219.
17. S. K. Smoukov, S. Gangwal, M. Marquez and O. D. Velev, *Soft Matter*, 2009, **5**, 1285-1292.
18. K. S. Khalil, A. Sagastegui, Y. Li, M. A. Tahir, J. E. Socolar, B. J. Wiley and B. B. Yellen, *Nature communications*, 2012, **3**, 794.
19. R. M. Erb, H. S. Son, B. Samanta, V. M. Rotello and B. B. Yellen, *Nature*, 2009, **457**, 999-1002.
20. Z. Ma, J. Guo, Y. Liu and Y. Ai, *Nanoscale*, 2015, **7**, 14047-14054.
21. Y. Chen, X. Ding, S. C. Lin, S. Yang, P. H. Huang, N. Nama, Y. Zhao, A. A. Nawaz, F. Guo, W. Wang, Y. Gu, T. E. Mallouk and T. J. Huang, *ACS Nano*, 2013, **7**, 3306-3314.
22. F. G. Mitri, F. H. Garzon and D. N. Sinha, *The Review of scientific instruments*, 2011, **82**, 034903.
23. B. Raeymaekers, C. Pantea and D. N. Sinha, *Journal of Applied Physics*, 2011, **109**, 014317.
24. M. Caleap and B. W. Drinkwater, *Proceedings of the National Academy of Sciences*, 2014, **111**, 6226-6230.
25. F. Xu, T. D. Finley, M. Turkaydin, Y. Sung, U. A. Gurkan, A. S. Yavuz, R. O. Guldiken and U. Demirci, *Biomaterials*, 2011, **32**, 7847-7855.
26. L. Schmid, D. A. Weitz and T. Franke, *Lab on a chip*, 2014, **14**, 3710-3718.
27. C. W. Shields IV, D. Sun, K. A. Johnson, K. A. Duval, A. V. Rodriguez, L. Gao, P. A. Dayton and G. P. Lopez, *Angew Chem Int Ed Engl*, 2014, **53**, 8070-8073.

28. M. S. Scholz, B. W. Drinkwater, T. M. Llewellyn-Jones and R. S. Trask, *IEEE Trans Ultrason Ferroelectr Freq Control*, 2015, **62**, 1845-1855.
29. F. Petersson, A. Nilsson, C. Holm, H. Jonsson and T. Laurell, *The Analyst*, 2004, **129**, 938-943.
30. X. Ding, S. C. Lin, B. Kiraly, H. Yue, S. Li, I. K. Chiang, J. Shi, S. J. Benkovic and T. J. Huang, *Proceedings of the National Academy of Sciences of the United States of America*, 2012, **109**, 11105-11109.
31. M. Antfolk, P. B. Muller, P. Augustsson, H. Bruus and T. Laurell, *Lab on a chip*, 2014, **14**, 2791-2799.
32. T. Laurell, F. Petersson and A. Nilsson, *Chemical Society Reviews*, 2007, **36**, 492-506.
33. M. Saito, K. Itagaki, K. Hayashi and K. Tsubata, *Japan Journal of Applied Physics*, 1999, **38**, 3028-3031.
34. M. Lu, S. Yang, Y. Ho, C. Grigsby, K. Leong and T. J. Huang, *ACS Nano*, 2014, **8**, 10026-10034.
35. G. R. Goddard, C. K. Sanders, J. C. Martin, G. Kaduchak and S. W. Graves, *Anal Chem*, 2007, **79**, 8740-8746.
36. M. E. Piyasena and S. W. Graves, *Lab on a chip*, 2014, **14**, 1044-1059.
37. L. M. Johnson, L. Gao, C. W. Shields IV, M. Smith, K. Efimenko, K. Cushing, J. Genzer and G. P. Lopez, *Journal of nanobiotechnology*, 2013, **11**, 22.
38. K. W. Cushing, M. E. Piyasena, N. J. Carroll, G. C. Maestas, B. A. Lopez, B. S. Edwards, S. W. Graves and G. P. Lopez, *Anal Chem*, 2013, **85**, 2208-2215.
39. F. Guo, W. Zhou, P. Li, Z. Mao, N. H. Yennawar, J. B. French and T. J. Huang, *Small*, 2015, **11**, 2733-2737.
40. J. Shi, H. Huang, Z. Stratton, Y. Huang and T. J. Huang, *Lab on a chip*, 2009, **9**, 3354-3359.
41. S. C. Lin, X. Mao and T. J. Huang, *Lab on a chip*, 2012, **12**, 2766-2770.
42. O. Jakobsson, C. Grenvall, M. Nordin, M. Evander and T. Laurell, *Lab on a chip*, 2014, **14**, 1943-1950.
43. C. W. Shields IV, C. D. Reyes and G. P. Lopez, *Lab on a chip*, 2015, **15**, 1230-1249.
44. L. Gao, C. W. Shields IV, L. M. Johnson, S. W. Graves, B. B. Yellen and G. P. Lopez, *Biomicrofluidics*, 2015, **9**, 014105.
45. C. W. Shields IV, L. M. Johnson, L. Gao and G. P. Lopez, *Langmuir*, 2014, **30**, 3923-3927.
46. M. Evander, L. Johansson, T. Lilliehorn, J. Piskur, M. Lindvall, S. Johansson, M. Almqvist, T. Laurell and J. Nilsson, *Anal Chem*, 2007, **79**, 2984-2991.
47. B. Hammarstrom, T. Laurell and J. Nilsson, *Lab on a chip*, 2012, **12**, 4296-4304.
48. B. Hammarstrom, B. Nilson, T. Laurell, J. Nilsson and S. Ekstrom, *Anal Chem*, 2014, **86**, 10560-10567.
49. R. R. Collino, T. R. Ray, R. C. Fleming, C. H. Sasaki, H. Haj-Hariri and M. R. Begley, *Extreme Mechanics Letters*, 2015, DOI: 10.1016/j.eml.2015.09.003.
50. Y. Yang, L. Gao, G. P. Lopez and B. B. Yellen, *ACS Nano*, 2013, **7**, 2705-2716.
51. Y. Yang, A. T. Pham, D. Cruz, C. Reyes, B. J. Wiley, G. P. Lopez and B. B. Yellen, *Advanced materials*, 2015, **27**, 4725-4731.
52. L. Gorkov, *Soviet-Physics-Doklady*, 1962, **6**, 773-775.
53. H. Bruus, *Lab on a chip*, 2012, **12**, 1014-1021.

54. D. T. Gillespie and E. Seitaridou, *Simple Brownian Diffusion: An Introduction to the Standard Theoretical Models*, Oxford University Press, Oxford, 2012.
55. R. Barnkob, P. Augustsson, T. Laurell and H. Bruus, *Lab on a chip*, 2010, **10**, 563-570.
56. J. Dual, P. Hahn, I. Leibacher, D. Moller, T. Schwarz and J. Wang, *Lab on a chip*, 2012, **12**, 4010-4021.
57. J. D. Weeks, D. Chandler and H. C. Anderson, *Journal of Chemical Physics*, 1970, **54**, 5237-5247.
58. D. Frenkel and B. Smit, *Understanding Molecular Simulation*, Academic Press, 2002.
59. F. G. Mitri and D. N. Sinha, *IEEE International Ultrasonics Symposium Proceedings*, 2011, 1556-1558.
60. C. R. Courtney, C. K. Ong, B. W. Drinkwater, P. D. Wilcox, C. Demore, S. Cochran, P. Glynne-Jones and M. Hill, *J Acoust Soc Am*, 2010, **128**, EL195-199.
61. C. R. P. Courtney, C. K. Ong, B. W. Drinkwater, A. L. Bernassau, P. D. Wilcox and D. R. S. Cumming, *Proceedings of the Royal Society A: Mathematical, Physical and Engineering Sciences*, 2011, **468**, 337-360.
62. J. Liu, C. W. Shields IV, O. Omofoye and G. P. Lopez, *MRS Proceedings*, 2014, **1622**, 55-60.
63. S. Oberti, A. Neild and J. Dual, *J. Acoust. Soc. Am.*, 2006, **121**, 778-786.
64. J. W. Swan, J. L. Bauer, Y. Liu and E. M. Furst, *Soft Matter*, 2014, **10**, 1102-1109.
65. S. M. Hagsater, T. G. Jensen, H. Bruus and J. P. Kutter, *Lab on a chip*, 2007, **7**, 1336-1344.
66. C. S. Kwiatkowski and P. L. Marston, *Journal of the Acousticial Society of America*, 1998, **103**, 3290-3300.
67. A. A. Doinikov, *J Fluid Mech*, 2001, **444**, 1-21.
68. J. Friend and L. Y. Yeo, *Reviews of Modern Physics*, 2011, **83**, 647-704.
69. A. G. Marin, H. Gelderblom, D. Lohse and J. H. Snoeijer, *Physical Review Letters*, 2011, **107**, 085502.
70. K. W. Wojciechowski and D. Frenkel, *Computational Methods in Science and Technology*, 2004, **10**, 235-255.
71. P.-Y. Wang, D. T. Bennetsen, M. Foss, T. Ameringer, H. Thissen and P. Kingshott, *Applied Materials & Interfaces*, 2015, **7**, 4979-4989.
72. G. Zhang and D. Wang, *Chemistry*, 2008, **4**, 236-245.
73. Y. Qiu, H. Wang, S. Gebhardt, A. Bolhovitins, C. E. Demore, A. Schonecker and S. Cochran, *Ultrasonics*, 2015, **62**, 136-146.
74. M. A. Bucaro, P. R. Kolodner, J. A. Taylor, A. Sidorenko, J. Aizenberg and T. N. Krupenkin, *Langmuir*, 2009, **25**, 3876-3879.
75. T. Wink, S. J. van Zuilen, A. Bult and W. P. van Bennekom, *The Analyst*, 1997, **122**, 43-50.
76. C. W. Shields IV, C. E. Livingston, B. B. Yellen, G. P. Lopez and D. M. Murdoch, *Biomicrofluidics*, 2014, **8**, 041101.
77. S. Li, F. Guo, Y. Chen, X. Ding, P. Li, L. Wang, C. E. Cameron and T. J. Huang, *Anal Chem*, 2014, **86**, 9853-9859.